\documentstyle[epsfig,natbib2,natbibmnfix]{mn2e}

\newcommand{\be}{\begin{equation}}
\newcommand{\ee}{\end{equation}}

\def\ltsima{$\; \buildrel < \over \sim \;$}
\def\simlt{\lower.5ex\hbox{\ltsima}}
\def\gtsima{$\; \buildrel > \over \sim \;$}
\def\simgt{\lower.5ex\hbox{\gtsima}}
\def\sgra{Sgr~A$^*$}

\def\msun{{\,{\rm M}_\odot}}

\def\del#1{{}}

\title{Accretion of cool stellar winds on \sgra: another puzzle
of the Galactic Centre?}

\author[J.~Cuadra, S.~Nayakshin, V.~Springel, \& T.~Di~Matteo]
{\parbox{18cm}{Jorge Cuadra$^{1}$\footnotemark[1], Sergei Nayakshin$^{1}$,
Volker Springel$^{1}$ \& Tiziana Di Matteo$^{1,2}$}\vspace{0.3cm}\\
$^1$Max-Planck-Institut f\"{u}r Astrophysik, Karl-Schwarzschild-Stra\ss{}e 1,
85741 Garching bei M\"{u}nchen, Germany\\ $^2$new adress: Department
of Physics, Carnegie Mellon University, Pittsburgh, PA 15213, USA}

\begin{document}

\maketitle

\begin{abstract}
\sgra\ is currently being fed by winds from a cluster of
gravitationally bound young mass-loosing stars. Using observational
constraints on the orbits, mass loss rates and wind velocities of
these stars, we numerically model the distribution of gas in the $\sim
0.1$--$10''$ region around \sgra. We find that radiative cooling of
recently discovered slow winds leads to the formation of many cool
filaments and blobs, and to a thin and rather light accretion disc of
about an arcsecond scale. The disc however does not extend all the way
to our inner boundary.  Instead, hot X-ray emitting gas dominates the
inner arcsecond. In our simulations, cool streams of gas frequently
enter this region on low angular momentum orbits, and are then
disrupted and heated up to the ambient hot gas temperature. The
accreting gas around \sgra\ is thus two-phase, with a hot component,
observable at X-ray wavelengths, and a cool component, which may be
responsible for the majority of time variability of \sgra\ emission on
hundred and thousand years time-scales. { We obtain an accretion
rate of a few $\times 10^{-6} \msun$~year$^{-1}$, consistent with {\em
Chandra} estimates, but variable on time-scales even shorter than
hundred years. These results strongly depend on the chosen stellar
orbits and wind parameters.  Further observational input is thus key
to a better modelling of \sgra\ wind accretion.}
\end{abstract}

\renewcommand{\thefootnote}{\fnsymbol{footnote}}
\footnotetext[1]{E-mail: {\tt jcuadra@mpa-garching.mpg.de}}

\section{Introduction}
\label{intro}

\sgra\ is identified with the $M_{\rm BH} \sim 3 \times 10^6 \msun$
super-massive black hole (SMBH) in the centre of our Galaxy
\citep[e.g., ][]{Schoedel02,Ghez03b}. By virtue of its location, {
\sgra\ may play a key role in the understanding of Active Galactic }
Nuclei (AGN).  Indeed, this is the only AGN where recent observations
detail the origin of the gas in the immediate vicinity of the SMBH
capture radius \citep[e.g.,][]{Najarro97, Paumard01, Baganoff03a,
Genzel03a}. This information, missing for all other AGN because of too
great a distance to them, their large luminosity, or both, is
absolutely necessary if the accretion problem is to be modelled
self-consistently.

Arguably the most famous puzzle of \sgra\ is its low luminosity with respect
to estimates of the accretion rate at around the capture radius, i.e.~at
distances of order $1'' \sim 10^5 R_{\rm S} \sim 0.04$ pc, where $R_{\rm S}$ is the
Schwarzschild radius of \sgra. Two methods have been deployed to obtain these
estimates. From {\em Chandra} observations of the Galactic centre region,
one can measure the gas density and temperature around the inner arcsecond and
then infer an estimate of the Bondi accretion rate of $\dot M \sim 10^{-6}
\msun$~year$^{-1}$\citep{Baganoff03a}.  However, unlike in the classical
textbook problem \citep{Bondi52}, hot gas is continuously created in shocked
winds expelled by tens of young massive stars near \sgra, and there is neither
a well defined concept of gas density and temperature at infinity, nor one for
the gas capture radius.

The other method addresses this problem by direct modelling of the gas
dynamics of stellar winds, assuming that the properties of the wind
sources are known. Three dimensional simulations of wind accretion
around \sgra\ were performed by \cite{Coker97},
{ 
who randomly positioned ten mass-loosing stars a few arc-seconds away
from \sgra. They presented two different runs in which the stars
were distributed in either a spherically isotropic or a flattened
system.
} 
\cite{Rockefeller04} used a particle-based code with more detailed
information on stellar coordinates and wind properties. However, in
both cases the stars were at fixed locations, whereas in reality they
follow Keplerian orbits around the SMBH.  
{ 
The accretion rate on to
\sgra\ predicted by both studies was estimated at $\sim \hbox{few}\;
\times 10^{-4} \msun$~year$^{-1}$. 
} 
Finally, \cite{Quataert04} studied
the problem in the approximation that there is an infinite number of
wind sources distributed isotropically around \sgra\, in a range of
radii. 
{ 
His model yields an accretion rate estimate of $\sim \hbox{few}\;
\times 10^{-5} \msun$~year$^{-1}$. 
}

Due to recent impressive progress in the observations of \sgra, we now know
much more about the origin of the gas feeding the SMBH.  Stellar wind sources
are locked into two rings that are roughly perpendicular to each other
\citep{Paumard01,Genzel03a}.  In addition, the wind velocities of several
important close stars were revised downward from $\sim 600$ km s$^{-1}$
\citep{Najarro97} to only $\sim 200$ km s$^{-1}$ \citep{Paumard01}, making the
Keplerian orbital motion much more important.

Motivated by these points, we performed numerical simulations of wind
accretion on to \sgra\ including optically thin radiative cooling and
allowing the wind-producing stars to be on Keplerian orbits. In this
Letter we report our most important findings: (i) the slow stellar
winds are susceptible to radiative cooling once they are shocked in
the hot bubble inflated by the fast winds; (ii) a disc/ring is formed
on $\sim$ arcsecond scale; (iii) cool gas blobs frequently enter the
inner arcsecond on low angular momentum orbits, are torn apart and
thermalized, and then mixed with the hot gas there. Gas is thus
distinctly two-phase in the inner region, with the cold phase being
invisible to {\em Chandra}. (iv) The resulting `accretion rate' (see
\S\ref{sec:results}) is of order several $\times 10^{-6}
\msun$~year$^{-1}$, consistent with {\em Chandra} estimates, but is
highly variable. This warrants a somewhat worrying question: how
representative is the current low luminosity of \sgra? Summarising
these points, it appears that the accretion flow on to \sgra\ and other
low luminosity AGN cannot be fully understood based on observations of
hot X-ray emitting gas alone.

\section{Analytical estimates}\label{sec:estimates}

The density of hot gas $1.5''$ away from \sgra\ is about $n_{\rm e} =
130$ cm$^{-3}$ \citep{Baganoff03a}, and the gas temperature is $T_{\rm
g} \approx 2\,$keV. The pressure of the hot gas is thus $p_{\rm th} =
n_{\rm e} T_{\rm g} \approx 3 \times 10^9$ K cm$^{-3}$. The
temperature resulting from collisions of stellar winds with wind
velocity $v_{\rm w} = 10^8 v_8$~cm~s$^{-1}$ is
\begin{equation}
T = 1.2 \times 10^7 \mu_{0.5} v_8^2 \quad \hbox{Kelvin},
\label{twind}
\end{equation}
where $\mu_{0.5}$ is the mean molecular weight in units of half the
proton mass. This temperature is to a large degree compatible with the
$T_{\rm g} \simeq 1.2\,$keV measured by {\em Chandra} at slightly
larger radii. The optically thin cooling function, dominated by metal
line emission, is $\Lambda \approx 6.0 \times 10^{-23} T_7^{-0.7}
(Z/3)$, where $T_7 = T/10^7$~K, and $Z$ is the metal abundance relative
to Solar \citep{Sutherland93}. The cooling time is thus
\begin{equation}
t_{\rm cool} = \frac{3 kT}{\Lambda n}\approx 10^4\;
\mu_{0.5}^{1.7} \; v_8^{5.4} \; p_*^{-1} \quad
\hbox{years}\;,
\end{equation}
where $p_* = nT/p_{\rm th}$. The dynamical time near this location is
$t_{\rm dyn} = R/v_{\rm K} \simeq 60 \, (R'')^{3/2}$ years, where
$R''$ is the radial distance to \sgra\ in arc-seconds
{ 
and $v_{\rm K}$ is the Keplerian velocity at that distance.
} 
This shows that cooling is of no importance for the gas originating in
the winds with outflow velocity $v_8\sim 1$.  However, if the wind
velocity is, say, $v_{\rm w}= 300$~km~s$^{-1}$, then $t_{\rm cool}$ is
only roughly 15 years, which is shorter than the dynamical
time. Therefore one could expect that slower winds may be susceptible
to radiative cooling in the high pressure environment of \sgra. In
reality the gas velocity is the sum of the wind velocity and the
stellar orbital motion. At a distance of $2''$, for example, the
Keplerian circular velocity is about 440 km~s$^{-1}$. With respect to
the ambient medium, the leading wind hemisphere will move with
velocity $v_8 \sim 1$, whereas the opposite one will move even slower
than $v_{\rm w}$. This shows that even for winds with velocities
$v_{\rm w} \simgt 500$ km s$^{-1}$, lagging regions of the wind may
still be affected by cooling.

\section{Method and initial conditions}\label{sec:method}

A full account of our numerical method along with  validation
tests will appear in \cite{Cuadra05}. Here we only briefly describe
the method.  We use the SPH/$N$-body code {\small GADGET-2}
\citep{Springel01,Springel02} to simulate the dynamics of
stars and gas in the gravitational field of the SMBH. This code,
developed for cosmological simulations, takes into account the
(Newtonian) {\em N}-body gravitational interactions of all particles
and also follows the hydrodynamics of the gas. We use the cooling
function cited in \S \ref{sec:estimates} with $Z=3$, and set the
minimum gas temperature to $10^4$ K.

We model the SMBH as a heavy `sink' particle \citep{Springel04,DiMatteo05},
with its mass set to $3.5 \times 10^6\msun$. For scales of interest, the
black hole gravity completely dominates over that of the surrounding
stars and gas. The inner boundary condition is specified by requiring
the gas passing within the radius $R_{\rm in}$ from the SMBH to
disappear in the black hole. In addition, particles at distances
larger than the `outer radius' $R_{\rm out}$ are of little interest
for our problem and are simply eliminated.

Stars are modelled as collisionless particles moving in the potential
of \sgra. The stars emit new gas particles that are initialised with
the minimum temperature
{ 
 and a mass $m_{\rm SPH} = 5\times10^{-7} \msun$.
} 
 The initial particle velocity is the sum of the orbital
motion of the star, and a random component. The latter is equal in
magnitude to the wind velocity and its direction is chosen randomly to
simulate isotropic winds in the frame moving with the star.

Following results of \cite{Paumard01}, we assume that `narrow line
stars' produce winds with velocity $v_{\rm w} = 300$ km s$^{-1}$, whereas
the `broad line stars' produce winds with $v_{\rm w} = 1000$ km s$^{-1}$.
We refer to the former as LBV stars, and to the latter as WR stars.
The radial extent of the inner stellar ring is from $2''$ to $5''$,
and the inner and outer radii of the outer ring are $4''$ and $8''$,
respectively. The rings are perpendicular to each other for simplicity
\citep[][conclude that the rings are inclined at $i\approx74^\circ$ to
each other]{Genzel03a}. We argue that the total number of wind sources
is likely to be higher than those that have been resolved so far. On
average they would clearly have to be less powerful than estimated by
\cite{Najarro97}. We therefore use 20 wind sources in total, with each
star loosing mass at the rate of $\dot M_* = 4 \times 10^{-5} \msun$
year$^{-1}$. Note that this is still a factor of $\sim 2-3$ below the
observationally estimated total mass loss rate from the \sgra\ star
cluster, but some of the wind sources are likely to be outside $R_{\rm
out}$ and thus should not be included here. 
{ 
Typically, the time-step of the calculation is $\sim$ 0.18 years, so
the mass
loss rate above implies that $\sim$ 20 SPH particles are created 
around each star per time-step. 
} 
The stars are distributed in rings uniformly in radius but randomly
along the azimuthal angle. Stars in the same ring rotate in the same
direction, of course, as they follow circular Keplerian orbits. We
populate the inner ring with 6 LBVs and 2 WRs, and the outer one with
3 LBVs and 9 WRs.

{ 
To increase the resolution in the inner region, we split the SPH
particles that get closer than $10\,R_{\rm in}$ to the SMBH. To
avoid numerical problems, the splitting is performed 
at a randomly chosen time - of the order of the dynamical time at that
radius - after the particle entered the inner $10\,R_{\rm in}$. Then 
the particle is divided into $N_{\rm split}=5$ new ones that
are placed randomly within the smoothing length of the original
one. The mass of the old particle is equally divided between the new
ones, while the temperature and velocity are kept constant. The rest
of the SPH properties are updated self-consistently by the code.  Other
simulations, without splitting, show the same overall results as the
one presented here \citep{Cuadra05}.
}

The values for the inner and outer boundary conditions are $R_{\rm in}
= 0.07''$ and $R_{\rm out} = 9''$. We start with a negligible gas
density inside the computational domain and then fill it up
self-consistently with stellar winds. We ran the simulation for about
4,000 years. Realistic longer simulations would have to include a
highly complex gas density field on larger scales around \sgra\ and
frequent supernovae (see below), which we leave for future
investigations. The total number of SPH particles, $N_{\rm SPH}$,
initially increases steeply with time until the hot winds fill up the
whole inner sphere. Later, the number of particles continues to
increase due to the fact that most of the slow wind particles are
bound.  By the end of the simulation we reach $N_{\rm SPH} = 1.7
\times 10^6$.

\section{Results}\label{sec:results}


\subsection{Gas morphology.} 
\begin{figure}
\centerline{\psfig{file=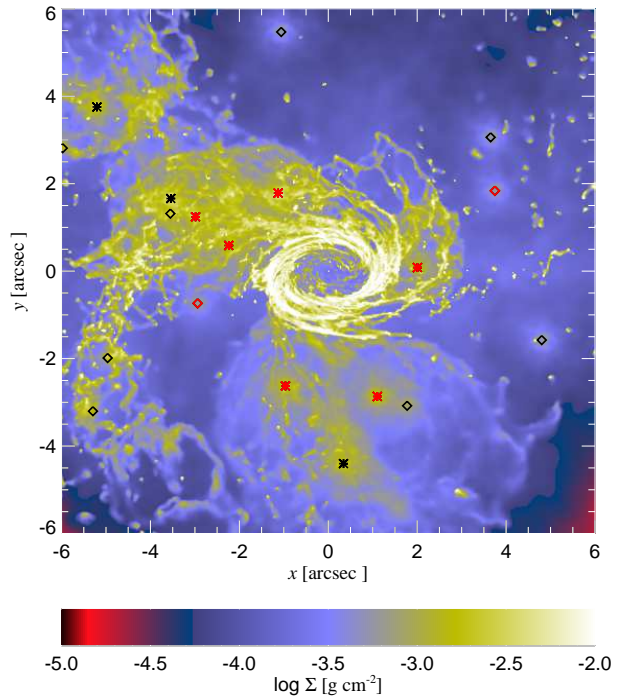,width=.5\textwidth,angle=0}}
\caption{View of the gas column density and of the stars (overlayed)
in the inner $6''$ cube of the computational domain, after about 3,000
years. The `LBV' stars within the slice are marked as 
{ 
stars,
while the `WR' stars are shown as diamonds. 
} 
Stars painted in
red and black belong to the inner and the outer rings, respectively.}
\label{fig:view}
\end{figure}

Figure \ref{fig:view} shows the column density of 
gas and the stellar wind sources in the inner $6''$ of the
computational domain. The inner and outer rings are viewed at
{ 
inclination angles of 40$^\circ$ and 50$^\circ$ in the Figure, respectively.
The inner ring rotates clockwise in this projection, while the outer
one rotates in the opposite sense.\footnote{An animated movie of this
simulation is available at\\ {\tt
http://www.mpa-garching.mpg.de/$\sim$jcuadra/Winds} ~.}
} 

As fast stellar winds fill the available space, the slower ones are
shocked and cool radiatively. Dense shells are formed around the
shocks. These shells are torn into filaments and blobs. Different
parts of the shells have different velocities and thus angular
momenta, creating many cool gas blobs with different velocities. Some
are directed to outer radii and have velocities large enough to escape
from the computational domain. Others receive velocities directing
them inward. At smaller radii the density of blobs is higher, thus
they collide, and settle in the plane established by the inner
LBVs. The blobs are also sheared by the differential Keplerian
rotation, and a gas disc is born at radii somewhat smaller than that
of the stellar orbits. 
{ 
The blobs typically consist of a few
hundred SPH particles, many more than the minimum 40 neighbours
used in the SPH kernel averaging for these simulations.
}

We find that the disc does not extend all the way inwards, as the
inner arcsecond or so is dominated by the hot X-ray emitting
gas. Surprisingly, the hot component originates from both fast and
slow winds. Cool filaments that enter the inner region appear to be
torn by differential rotation and then shock-heated to the ambient
temperature as they interact with the hot gas.  Since the disc mass is
miniscule by AGN standards, the disc is constantly violently affected
by the stars from the other ring. Streams and blobs of cool gas are
kicked high up above the disc midplane due to interactions with winds
from stars crossing the disc. These blobs rain down on to the disc, and
some are brought into the sub-arcsecond region of the flow.

\subsection{Accretion on to \sgra.} 
\begin{figure}
\centerline{\psfig{file=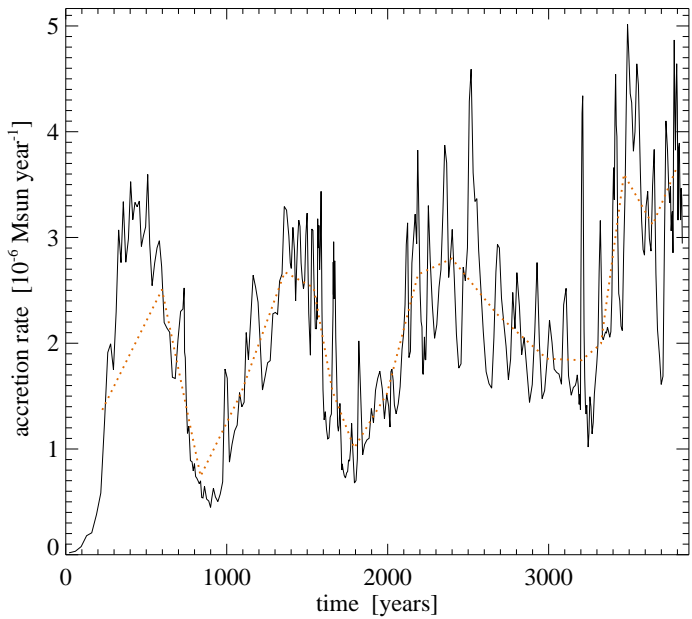,width=.5\textwidth,angle=90}}
\caption{Accretion rate on to \sgra\ versus time, averaged in bins of
10 years (solid) and 200 years (dotted). Both are expected to be only
rough indicators of the actual accretion rate since much of the gas
captured may still be unbounded inside $R_{\rm in}$ and/or has to be
processed in the inner accretion flow that we do not resolve.}
\label{fig:accrate}
\end{figure}

The `accretion rate', $\dot{M}_{\rm BH}$, 
 on to \sgra\ is defined in our simulations as the rate at which SPH
particles enter the sphere with radius $R_{\rm in}$, and is plotted
versus time in Fig. \ref{fig:accrate}. Solid and dotted lines
correspond to time bins of 10 and 200 years, respectively. The latter
seems to be a reasonable estimate for the viscous time-scale of the
hot accretion flow at around $R_{\rm in}$. 
{
The average rate we obtain, 
a few $\times 10^{-6} \msun$~year$^{-1}$, is in good agreement with
the Bondi estimate \citep{Baganoff03a}, and is one to two orders of magnitude
lower than what previous studies found (cf. \S\ref{intro}).
However, factors 
} 
of a few
variability in the accretion rate are immediately obvious. This
variability can be tracked down to the arrival of individual cool gas
blobs or filaments in the innermost region.  Since the density of the
hot gas in the vicinity of \sgra\ is never zero, the accretion rate
never decreases to zero.

It must be stressed that the real $\dot{M}_{\rm BH}$ further depends
on intricate physical details of the inner accretion flow that we
cannot resolve here. Some of the gas entering $R \leq R_{\rm in}$ is
unbound and some may become unbound later on as result of viscous
heating in the flow \citep{Blandford99}.  Therefore, the accretion
rate measured in our simulations is best understood to yield the outer
boundary conditions for the inner accretion flow, and as such should
be a more physically complete estimate of that than the commonly used
Bondi accretion rate based on {\em Chandra} observations of hot X-ray
emitting gas only.

{ 
The number of particles in the inner $1''$ (comparable to the Bondi
radius estimate for this problem) is about 70,000. This ensures that
we have enough resolution at the inner boundary. On average, $\sim$ 4 SPH
particles are accreted at each time-step.
} 

\subsection{Existence of cold disc in sub-arcsecond region of \sgra.}

The cold `disc' found in our simulations is constantly being created
but it is also likely to be destroyed from time to time. Its mass can
be estimated as $M_{\rm disc} = \dot{M}_{\rm cw} t = 10 \msun
\;(\dot{M}_{\rm cw}/10^{-3} \msun \hbox{year}^{-1})\; t_4$, where $t_4
= t/10^4$~years and $\dot{M}_{\rm cw}$ is the mass outflow rate of
cool stellar winds. A supernova occurring within the inner $0.5$~pc of
the Galaxy would easily destroy such a disc. The number of young early
type stars in the \sgra\ cluster is likely to be in the hundreds, and
some have already reached the WR stage
\citep[e.g.][]{Paumard01,Genzel03a}.  Assuming that the most recent
star formation event occurred a few million years ago, one would then
estimate the SN rate in \sgra's inner star cluster to be at least $\sim
10^{-4}$ per year. For comparison, the viscous time at a radius of $1''$
is $10^6-10^7$ years at best \citep[e.g. fig. 2 in][]{NC05}.
Furthermore, there is much more cold material on scales of several
arcseconds and beyond that may be plunging on to \sgra
\citep{Paumard04}. Arrival of this mass in the inner arcsecond could
also destroy the disc.

Therefore it is likely that the cool disc of $\sim$ arcsecond scales
does not smoothly extend inside the sub-arcsecond region of
\sgra. However, occasionally we observe cool gas blobs to directly
fall into the capture `sphere' in the simulation. It is also
conceivable that events started by a supernova shell passage, or by an
infall of additional cool material, could also leave a remnant in the
form of a cold disc.  Thermal conduction between the two phases could
lead to evaporation of a cold accretion disc via the \cite{Meyer94}
mechanism, which is unfortunately model dependent due to our poor knowledge
of the magnetic field geometry and strength.

\cite{NS03} suggested, mainly based on the presence of X-ray flares in
\sgra, that there is a cool disc at $\sim 0.01$--$0.1''$ scales and
beyond. The required mass of the disc was estimated to be smaller than
a Solar mass. However, observations of NIR flares \citep{Genzel03b}
put the flares at no more than a few milli-arcseconds away from the
radio position of \sgra, which is somewhat problematic for this
model. Also, the NIR flare spectra appear to strongly favour
a synchrotron (i.e. SMBH jet) origin. Further, no eclipses or star
`brightenings' expected when bright stars approach the disc
\citep{Cuadra04} have been observed so far. Summarising, there is currently
no observational motivation to favour the existence of such a disc in
\sgra, although it is also difficult to rule out its presence
\citep[unless the disc extends to the SMBH horizon;][]{Falcke97}.

Concluding, we suggest that there cannot be a smooth transition of the
larger scale cool disc into the inner sub-arcsecond regions of
\sgra. Nevertheless, the issue of the current existence of a cool disc
or its periodic appearance and disappearance is an open subject for
future work.

{

\subsection{Note on the importance of initial conditions.}

While this work appears to be the most detailed (to date) numerical
attempt to model the accretion of stellar winds on to \sgra,
observational uncertainties in the stellar orbits and wind mass loss
rates and velocities still leave a lot of room for uncertainties in
the final results. The latest observations (R. Genzel \& T. Paumard,
priv. comm.) reveal that the narrow emission line stars might be more
equally divided between the two discs than what we have assumed here.
In this case, the disc-like structure would form at a larger scale and
be probably not as conspicuous. Similarly, if the mass loss rate of
the `LBV' stars is smaller, the cool disc becomes obviously less
massive. In addition, mass-loss rates of LBV stars have been
observed to vary by more than an order of magnitude within a few years
\citep[e.g.,][]{Leitherer97}. This effect would
bring further variability and uncertainty to the results.

Another important ingredient, still missing in our approach, is the
inclusion of cooler gas filaments observed to be (possibly) infalling
on to \sgra\ on several to tens of arcsecond scales from \sgra.  These
structures, referred to as the `mini-spiral' \citep{Scoville03,
Paumard04}, would undoubtfully change some of our results.

Finally, like all previous numerical investigations
\citep{Coker97,Rockefeller04,Quataert04}, we have entirely neglected
the contribution of the wind producing stars on orbits inside the
central $2''$. While the most important wind sources are all located
outside of this region \citep{Najarro97,Paumard01}, the weaker OB
stellar winds inside the inner region may still contribute. For
example, if their wind loss rates are about the inferred accretion
rate, these stars \citep[that appear to be on more randomly oriented
orbits, see][]{Eisenhauer05} could potentially destroy the disc in the
inner arcsecond.\footnote{Note that an extreme case of this was already
studied analytically assuming a ballistic trajectory approximation by
\cite{Loeb04}. He showed that if the inner `S'-stars
\citep[e.g.,][]{Ghez05,Eisenhauer05} produce wind outflow rates as
strong as $\sim 10^{-6} \msun$~year$^{-1}$, then their winds alone
could provide enough fuel for \sgra\ emission. However, most of the
`S'-stars now appear to be of intermediate to later B-type, suggesting
their winds could be orders of magnitude weaker than assumed by
\cite{Loeb04}.}

Future observations of the wind properties and stellar orbits near
\sgra\ are key to produce an increasingly more realistic model of the
accretion flow on to \sgra.}

\section{Discussion and Conclusions}\label{sec:conclusions}

AGN accretion discs are less well understood than discs in X-ray
binaries on comparable relative scales, because we do not have good
observational constraints on the origin of gas accreting on to the
SMBH. \sgra\ is becoming the only exception to this as observations
improve. In this paper we made an attempt to realistically simulate
the outer $\sim 0.1$--$10''$ region of the gas flow on to \sgra. The
resulting gas flow is far more complex than thought earlier based on
studies that included non-radiative fast stellar winds from stars
either fixed in space or distributed in a spherically-symmetric
fashion.  The presence of cool gas in the sub-arcsecond region, as
found here, may considerably complicate the interpretation of
observational constraints on the accretion of \sgra. { Although, as
discussed above, this does depend on still somewhat poorly known details
of stellar orbits and wind parameters.}

{ The average accretion rate in our simulation, a few $\times
10^{-6} \msun$~year$^{-1}$, is consistent with the estimates of
\cite{Baganoff03a} and 1--2 orders of magnitude lower than what previous
models found \citep{Coker97,Rockefeller04,Quataert04}. However, this
accretion rate changes by factors of a few in time-scales shorter than
hundred years. Then how } representative is the current low luminosity
state of \sgra, if the feeding of the inner region is indeed so
turbulent and time-variable as our simulation suggests? { Afterall,
observations of X-ray/$\gamma$-ray echoes from nearby molecular clouds
indicate that \sgra\ might have been much more luminous some $\sim
300$ years ago \citep{Sunyaev93,Koyama96,Revnivtsev04}.}  Another
aspect of the same issue is that `accretion' of a cool blob in our
simulations is not yet a true accretion event. If the blob manages to
survive in the hot gas and settles to a disc or a ring at say $\sim
10^3 R_{\rm S}$, it may circle \sgra\ for a long time without being
noticed. Further uncertainty in these results is the interaction
between the hot and the cold gas via thermal conduction. If a cold
blob enters the inner region of the hot flow, and is evaporated there,
how will this affect the flow there? These and other related questions
are to be resolved in future work if we want to reach a full
understanding of the accretion process on to \sgra.

\bibliographystyle{mnras}
\bibliography{nayakshin}

\end{document}